\documentclass[11pt]{article}
\usepackage{erice,epsfig}

\def\be{\begin{equation}}
\def\ee{\end{equation}}
\def\bea{\begin{eqnarray}}
\def\eea{\end{eqnarray}}

\begin{document}
%\vspace*{4cm}

\begin{flushright}
UCLA/02/TEP/5
\end{flushright}

\title{Future determination of the neutrino-nucleon cross section at
extreme energies.\footnote{Talk presented at {\em Workshop on electromagnetic
probes of fundamental physics}, October 16-21, 2001, Erice, Italy} }

\author{ Alexander Kusenko }

\address{Department of Physics and Astronomy, UCLA, Los Angeles, CA
90095-1547\\ RIKEN BNL Research Center, Brookhaven National
Laboratory, Upton, NY 11973 
}

\maketitle

\abstracts{ 
Future detectors of cosmic rays, such as EUSO and
OWL, can test the Standard Model predictions for the neutrino interactions
at energies well beyond the reach of any terrestrial experiment.  The
relative rates of horizontal and upgoing air showers, combined with the
angular distribution of upgoing air showers will allow one to measure the
neutrino-nucleon cross section at $\sqrt{s} \sim 10^5$~GeV or higher.}

Detection of ultrahigh-energy (UHE) neutrinos will mark the advent of UHE
neutrino astronomy, allowing the mapping on the sky of the most energetic
and most distant sources in the Universe.  In addition, the prospects for
detection of the neutrino-induced upgoing air showers (UAS) by the
ground-level fluorescence detectors~\cite{Feng}, such as HiRes, Telescope
Array, and Pierre Auger~\cite{Bertou}, as well as the future orbiting
detectors~\cite{Domokos,Fargion}, present an opportunity to conduct a {\em
particle physics experiment}~\cite{kw} and measure the neutrino cross section
$\sigma_{\nu_N}$ at an unprecedented center-of-mass energy $10^5-10^6$~GeV.
The relative rates of the horizontal air showers (HAS) and UAS initiated by
neutrinos depend on $\sigma_{\nu_N}$ in such a way that the cross section
can be determined without a precise knowledge of the incident neutrino
flux.  Moreover, the angular distribution of UAS provides an additional and
independent information about the cross section.

The first question is, of course, whether there is a sufficient flux of
neutrinos to detect.  Observations of ultrahigh-energy cosmic rays (UHECR)
imply the existence of a related flux of ultrahigh-energy neutrinos
generated in the interactions of UHECR with cosmic microwave background
radiation~\cite{UHECR}.  In addition, active galactic nuclei~\cite{AGN},
gamma-ray bursts~\cite{GRB}, and other astrophysical objects can produce a
large flux of neutrinos~\cite{Waxman}.  Finally, if the solution to the
puzzle~\cite{UHECR} of UHECR involves Z-bursts~\cite{Zburst}, there is a
strong additional flux of ultrahigh-energy neutrinos~\cite{Zburst1,Zburst2}.
The flux of UHE neutrinos at energies $10^{18}-10^{20}$~eV is uncertain.
However, as discussed below, the proposed measurement of the neutrino cross
section is not sensitive to these uncertainties~\cite{kw}.

Calculations of the neutrino-nucleon cross section $\sigma_{\nu_N}$ at
$10^{20}$~eV necessarily use an extrapolation of parton distribution
functions and Standard Model parameters far beyond the reach of present
experimental data.  The resulting cross section~\cite{UHEnusig} at
$10^{20}$~eV is $\sim 10^{-31}{\rm cm}^2$.  It is of great interest to
compare this prediction with experiment to test the small-$x$ behavior of
QCD, as well as the possible contributions of new physics beyond the
electroweak scale.

For the purposes of such a measurement, we assume the cross
section to be a free parameter bounded from below by the value $\sim
2\times 10^{-34}{\rm cm}^2$ measured at HERA at $\sqrt{s}=314$~GeV. (This 
corresponds to a laboratory energy $E_\nu=5.2\times 10^{13}$~eV of an
incident neutrino.)

UHE neutrinos are expected to arise from pion and muon decays.  The subsequent
oscillations generate a roughly equal fraction of each neutrino flavor. Tau
neutrinos interacting below the surface of the Earth can create an
energetic $\tau$-lepton, whose decay in the atmosphere produces an UAS.

It is clear that, for smaller values of the cross section, the Earth is
more transparent for neutrinos, so that more of them can interact just
below the surface and produce a $\tau$ that can come out into the
atmosphere.  As long as the mean free path $\lambda_\nu $ is smaller than
the radius of the Earth, the rates of UAS increase with $\lambda_\nu
\propto 1/\sigma_{\nu_N} $.  The rates of HAS, however, are proportional to
$\sigma_{\nu_N}$; they decrease for a smaller cross section.  The
comparison of the two rates, shown in Fig.~1, can allow a measurement of
the cross section which is practically independent of the uncertainties in
the incident neutrino flux.

\begin{figure}[t]
\centering
\hspace*{-5.5mm}
\leavevmode\epsfysize=6cm %\epsfxsize=8cm 
\epsfbox{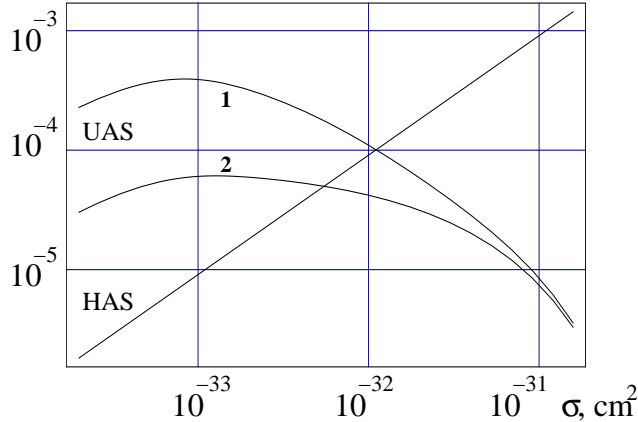} 
\caption[fig. 2]{\label{fig2} The air shower probability per incident tau
neutrino as a function of the neutrino
cross section. 
The incident neutrino energy is $10^{20}$~eV and the assumed 
energy threshold for detection of UAS is
$E_{\rm th}=10^{18}$eV for curve 1 and $ 10^{19}$eV for curve 2.
}
\end{figure} 

In addition, the angular distribution of UAS alone can be used as an
independent measurement of the cross section.  The peak of the angular
distribution of UAS occurs~\cite{kw} when $\cos \theta_{\rm peak} \approx 
\lambda_\nu/ 2 R_\oplus$, which depends on the cross section. 

It is comforting to know that the program of UHE neutrino astronomy, which
is one of the goals of EUSO and OWL, is not at risk, regardless of any
theoretical uncertainties in the neutrino cross section.  For a larger
cross section, HAS are more frequent than HAS, while for a smaller value
UAS dominate.  Nevertheless, the total rates of combined events remain
roughly constant for a wide range of $\sigma_{\nu_N}$, as shown in Fig.~1.

On the other hand, some of the reported bounds on the neutrino flux are
directly affected by the uncertainties in the neutrino-nucleon cross
section.  For example, the reported bounds on the UHE neutrino flux due to
the non-observation of neutrino-initiated HAS~\cite{FlyEye} and of radio
signals produced by neutrino interactions near the surface of the
moon~\cite{GLN99} are weaker if the cross section is smaller.

To conclude, the future neutrino cosmic-ray experiments can determine the
neutrino-nucleon cross section at energies as high as $10^{11}$~GeV, or
higher, by comparing the rates of UAS with those of HAS; or by measuring
the angular distribution of UAS events.  Hence, there is an exciting
opportunity do a particle physics experiment using a natural ``beam'' of
cosmic UHE neutrinos in the near future.  In addition, the overall
prospects for UHE neutrino astronomy are not marred by possible theoretical
uncertainties in the value of the neutrino-nucleon cross section: the total
number of horizontal and upgoing events remains sufficient for a wide range
of $\sigma_{\nu_N}$.

\section*{Acknowledgments}

This work was supported in part by the DOE grant DE-FG03-91ER40662.


\section*{References}
\begin{thebibliography}{99}

\bibitem{Feng}
J.~L.~Feng, P.~Fisher, F.~Wilczek and T.~M.~Yu,
%``Observability of earth-skimming ultra-high energy neutrinos,''
arXiv:hep-ph/0105067.
%%CITATION = HEP-PH 0105067;%%

\bibitem{Bertou}
X.~Bertou, P.~Billoir, O.~Deligny, C.~Lachaud and A.~Letessier-Selvon,
%``Tau neutrinos in the Auger observatory: A new window to UHECR sources,''
arXiv:astro-ph/0104452.
%%CITATION = ASTRO-PH 0104452;%%

\bibitem{Domokos}
G.~Domokos and S.~Kovesi-Domokos,
%``Observation of ultrahigh energy neutrino interactions by orbiting  detectors,''
arXiv:hep-ph/9805221.
%%CITATION = HEP-PH 9805221;%%

\bibitem{Fargion}
D.~Fargion,
%``Discovering ultra high energy neutrinos by horizontal and upward tau
%air-showers: First evidences in terrestrial gamma flashes,'' 
arXiv:astro-ph/0002453.
%%CITATION = ASTRO-PH 0002453;%%

\bibitem{kw}
A.~Kusenko and T.~Weiler,
%``Neutrino cross sections at high energies and the future observations of
%ultrahigh-energy cosmic rays,'' 
arXiv:hep-ph/0106071.
%%CITATION = HEP-PH 0106071;%%

\bibitem{UHECR}
For reviews, see, {\em e.g.}, 
P. Bhattacharjee and G. Sigl, Phys. Rept. 327, 109 (2000);
M. Nagano and A. A. Watson,
%"Observations and implications of the ultrahigh-energy cosmic rays,''
Rev. Mod. Phys. 72, 689 (2000);
T.~J.~Weiler,
%``Extreme-energy cosmic rays: Puzzles, models, and maybe neutrinos,''
arXiv:hep-ph/0103023; 
%%CITATION = HEP-PH 0103023;%%
P.~Biermann and G.~Sigl,
%``Introduction to Cosmic Rays,''
arXiv:astro-ph/0202425.
%%CITATION = ASTRO-PH 0202425;%%

\bibitem{AGN}
F.~W.~Stecker, C.~Done, M.~H.~Salamon and P.~Sommers,
%``High energy neutrinos from the cores and jets of active galaxies,''
{\it Given at High-energy Neutrino Astrophysics Workshop on Astrophysics of
High-energy Neutrinos: Particle Physics, Sources, Production Mechanisms and
Detection Prospects, Honolulu, Hawaii, 23-26 Mar 1992}; 
K.~Mannheim,
%``High-Energy Neutrinos From Extragalactic Jets,''
Astropart.\ Phys.\  {\bf 3}, 295 (1995); 
%%CITATION = APHYE,3,295;%%
R.~J.~Protheroe,
%``High energy neutrinos from blazars,''
arXiv:astro-ph/9607165; 
%%CITATION = ASTRO-PH 9607165;%%
F.~Halzen and E.~Zas,
%``Neutrino fluxes from active galaxies: A model-independent estimate,''
Astrophys.\ J.\  {\bf 488}, 669 (1997)
[arXiv:astro-ph/9702193].
%%CITATION = ASTRO-PH 9702193;%%

\bibitem{GRB} E.~Waxman,
%``High energy cosmic-rays and neutrinos from cosmological gamma-ray burst
%fireballs,'' 
Phys.\ Scripta {\bf T85}, 117 (2000)
[arXiv:astro-ph/9911395]; 
%%CITATION = ASTRO-PH 9911395;%%
%``Gamma-ray bursts, cosmic-rays and neutrinos,''
Nucl.\ Phys.\ Proc.\ Suppl.\  {\bf 87}, 345 (2000)
[arXiv:astro-ph/0002243].
%%CITATION = ASTRO-PH 0002243;%%

\bibitem{Zburst} T.~Weiler, Phys. Rev. Lett. 49, 234 (1982); 
Astropart. Phys. {\bf 11}, 303 (1999); 
%[hep-ph/9710431].   
%%CITATION = HEP-PH 9710431;%%
D.~Fargion, B.~Mele and A.~Salis,
%``Ultrahigh energy neutrino scattering onto relic light neutrinos in
%galactic halo as a possible source of highest energy extragalactic  cosmic
%rays,''
Astrophys.\ J.\  {\bf 517}, 725 (1999). 

\bibitem{Waxman}
For a recent review, see, {\it e.g.}, E.~Waxman,
%``High Energy Neutrinos From Astrophysical Sources,''
Nucl.\ Phys.\ Proc.\ Suppl.\  {\bf 100}, 314 (2001).
%%CITATION = NUPHZ,100,314;%%

\bibitem{Zburst1}
R.J. Protheroe, astro-ph/9809144; 
G.~Gelmini and A.~Kusenko,
%``Highest-energy cosmic rays from Fermi-degenerate relic neutrinos
%consistent with Super-Kamiokande results,'' 
Phys.\ Rev.\ Lett.\  {\bf 82}, 5202 (1999); 
%[hep-ph/9902354].
%%CITATION = HEP-PH 9902354;%%
%``Unstable superheavy relic particles as a source of neutrinos  responsible for the ultrahigh-energy cosmic rays,''
Phys.\ Rev.\ Lett.\  {\bf 84}, 1378 (2000);
%[arXiv:hep-ph/9908276].
%%CITATION = HEP-PH 9908276;%%
S. Yoshida, G. Sigl and S.-J. Lee, 
Phys. Rev. Lett. 81, 5505, (1998);
%"Extremely high energy neutrinos, neutrino hot dark matter,
%and the  highest energy cosmic rays",
%[hep-ph/9808324]
%%CITATION = HEP-PH 9808324;%%
K. Mannheim, R.J. Protheroe, J.P. Rachen, Phys.\ Rev.\ D63, 023003 (2001); 
J.L. Crooks, J.O. Dunn, and P.H. Frampton, astro-ph/0002089;
Z. Fodor, S. D. Katz and A. Ringwald, [hep-ph/0105064]; 
%"Determination of absolute neutrino masses from Z-bursts",
%%CITATION = HEP-PH 0105064;%%
%

\bibitem{Zburst2}
G.~B.~Gelmini,
%``Super-Kamiokande 0.07-eV neutrinos in cosmology: Hot dark matter and
%the highest energy cosmic rays,'' 
hep-ph/0005263;
%%CITATION = HEP-PH 0005263;%%
J.~Alvarez-Muniz and F.~Halzen,
%``10**20-eV cosmic ray and particle physics with IceCube,''
arXiv:astro-ph/0102106; 
%%CITATION = ASTRO-PH 0102106;%%
O.~E.~Kalashev, V.~A.~Kuzmin, D.~V.~Semikoz and G.~Sigl,
%``Ultra-high energy cosmic rays from neutrino emitting acceleration
%sources?,'' 
arXiv:hep-ph/0112351; 
%%CITATION = HEP-PH 0112351;%%
G.~Gelmini and G.~Varieschi,
%``Cosmic rays above the ankle from Z-bursts with 0.07-eV relic neutrinos,''
%arXiv:
hep-ph/0201273.
%%CITATION = HEP-PH 0201273;%%

\bibitem{UHEnusig}
  G.M. Frichter, D.W. McKay and J.P. Ralston,
      { Phys. Rev. Lett.} {\bf 74}, 1508 (1995);
      { Phys. Rev. Lett.} {\bf 77E}, 4107 (1996);
R.~Gandhi, C.~Quigg, M.~H.~Reno and I.~Sarcevic,
%``Ultrahigh-energy neutrino interactions,''
Astropart.\ Phys.\  {\bf 5}, 81 (1996); 
%[hep-ph/9512364].
%%CITATION = HEP-PH 9512364;%%
R.~Gandhi, C.~Quigg, M.~H.~Reno and I.~Sarcevic,
%``Neutrino interactions at ultrahigh energies,''
Phys.\ Rev.\ D {\bf 58}, 093009 (1998)
%[hep-ph/9807264].
%%CITATION = HEP-PH 9807264;%%

\bibitem{FlyEye}
R.M. Baltrusaitis, et al.\ (Fly's Eye Collaboration),
Phys. Rev. D31, 2192 (1985).

\bibitem{GLN99}  P. Gorham, K. Liewer, and C. Naudet,
  [astro-ph/9906504] and Proc.\ 26th Int. CR Conf., 
   Salt Lake City, Utah, Aug. 1999; 
~Alvarez-Muniz and E.~Zas,
%``Prospects for radio detection of extremely high energy cosmic rays and
%neutrinos in the moon,'' 
astro-ph/0102173; 
%%CITATION = ASTRO-PH 0102173;%%
P.~W.~Gorham, K.~M.~Liewer, C.~J.~Naudet, D.~P.~Saltzberg and D.~R.~Williams,
%``Radio limits on an isotropic flux of >100 EeV cosmic neutrinos,''
astro-ph/0102435.
%%CITATION = ASTRO-PH 0102435;%%


\end{thebibliography}
\end{document}